\documentclass[reqno, 12pt]{amsart}

\usepackage{amsmath,amssymb,braket,appendix,nicefrac,enumitem,upgreek,dsfont}
\usepackage[protrusion=true,babel=true]{microtype}
\usepackage[english]{babel}
\usepackage[widespace]{fourier}
\usepackage[backrefs]{amsrefs}
\usepackage[margin=1in]{geometry}
\usepackage[onehalfspacing]{setspace}
\usepackage[pdfusetitle, pagebackref]{hyperref}

\numberwithin{equation}{section}

\usepackage[nameinlink, noabbrev]{cleveref}
\expandafter\def\csname ver@etex.sty\endcsname{3000/12/31}

\usepackage{autonum}

\allowdisplaybreaks[1]

\let\originalleft\left
\let\originalright\right
\renewcommand{\left}{\mathopen{}\mathclose\bgroup\originalleft}
\renewcommand{\right}{\aftergroup\egroup\originalright}

\def\({\mathopen{}\left(}
\def\){\right)\mathclose{}}

\makeatletter
\renewcommand*{\eqref}[1]{\hyperref[{#1}]{\textup{\tagform@{\ref*{#1}}}}}
\makeatother

\newcommand*{\eqdef}{\mathrel{\vcenter{\baselineskip0.5ex \lineskiplimit0pt\hbox{.}\hbox{.}}}=}

\newtheorem{theorem}{Theorem}[section]
\newtheorem{lemma}[theorem]{Lemma}
\newtheorem{corollary}[theorem]{Corollary}
\newtheorem{remark}[theorem]{Remark}

\crefname{theorem}{Theorem}{Theorems}
\creflabelformat{theorem}{#2{#1}#3}
\crefname{lemma}{Lemma}{Lemmas}
\creflabelformat{lemma}{#2{#1}#3}
\crefname{corollary}{Corollary}{Corollaries}
\creflabelformat{corollary}{#2{#1}#3}
\crefname{remark}{Remark}{Remarks}
\creflabelformat{remark}{#2{#1}#3}

\crefname{ineq}{inequality}{inequalities}
\creflabelformat{ineq}{#2{\upshape(#1)}#3}

\def\id{\mathds{1}}
\def\cx{\mathbb{C}}
\def\F{\mathbb{F}}

\def\N{\mathbb{N}}
\def\rl{\mathbb{R}}

\def\Z{\mathbb{Z}}

\def\CNOT{\mathrm{CNOT}}
\def\supp{\mathrm{supp}}

\def\QROM{\mathrm{QROM}}
\def\cR{\mathcal{R}}

\def\cW{\mathcal{W}}
\def\wh{\mathrm{wh}}

\title{Shallow data-loading circuits for multivariable functions}
\date{\today}

\author{Arjun Dixit}
\address[Arjun Dixit]{Department of Physics, University of Toronto, Toronto, Ontario, Canada}
\email{\href{mailto:arjun.adixit@mail.utoronto.ca}{arjun.adixit@mail.utoronto.ca}}

\author{\'Akos Nagy}
\address[\'Akos Nagy]{Toronto, Ontario, Canada}
\email{\href{mailto:contact@akosnagy.com}{contact@akosnagy.com}}
\urladdr{\href{https://akosnagy.com/}{akosnagy.com}}

\hypersetup{
    unicode         = true,
    pdffitwindow    = true,
    pdftoolbar      = false,
    pdfmenubar      = false,
    pdfstartview    = {FitH},
    hypertexnames   = true,
    colorlinks      = true,
    linkcolor       = black,
    citecolor       = black,
    filecolor       = black,
    urlcolor        = blue
}

\calclayout
\begin{document}

\begin{abstract}
    We construct shallow QROM, diagonal block-encoding, and quantum state preparation oracles for analytic functions on the $D$-dimensional hypercube $[0, 1]^D$. On $n$ qubits and with $2^{- \Omega \( \nicefrac{n}{D} \)}$ error, these circuits have depths of $O \( \sqrt{\tfrac{n \log_2 (n)}{D}} \)$ in the Clifford + $T$ gate set, albeit using $2^{O(\sqrt{nD \log_2 (n)})}$ ancilla qubits. This is a near-quadratic depth reduction compared to known linear-depth methods.
\end{abstract}

\maketitle

\section{Introduction}
\label{sec:intro}

Efficient data-loading methods are central to many quantum algorithms. In this paper we consider three kinds of data-loading primitives: QROMs, block-encodings of diagonal operators, and quantum state preparation. More concretely, we study the problem when the inputs are given by multivariable functions. This problem has been studied extensively, especially for quantum simulations and PDEs; cf. \cites{Holmes2020SmoothStatePreparation,MarinSanchez2023ApproximateFunctionLoading,Motlagh2024VibronicDynamics,yano2026quantum,Sturm2026ExplicitBlockEncodingLaplacian,zhao2021smooth,guseynov2026efficient,moosa2024linear,zylberman2025efficient,o2025quantum}.

In \cite{zylberman2025efficient}*{Theorem IV.1} the authors make an important point about encoding differentiable functions, which was generalized in \cite{szczepanik2026simulating}*{Theorem A.1}, and further extended below in \Cref{lemma:smooth_encoding_scaling}: if the function is ``smooth enough'' (differentiable or at least Lipschitz), then there is a natural correspondence between the number of qubits (corresponding to discretization error) and the error of the block-encodings and quantum state preparation circuits. Using unnecessarily high precision wastes gates, while using unnecessarily low precision wastes qubits. Assuming that all dimensions are discretized with $m$ qubits (thus, the total number of qubits is $n = D m$, where $D$ is the number of variables), this error is $O \( 2^{- m} \)$ for both block-encoding and quantum state preparation.

The depths of the known methods scale at least linearly with the number of qubits; for example, \cite{moosa2024linear} presents a linear-depth method based on Fourier series approximations of the function. Our goal in this paper is to present shallow implementations of such quantum subroutines, achieving a near-quadratic depth reduction; more precisely, the depth of the quantum state preparation circuit of such an $n$-qubit state scales as $O \( \sqrt{\tfrac{n \log_2 (n)}{D}} \)$, where we omitted constant factors depending only on the analytic properties of the input function; cf. \Cref{theorem:sp_analytic}. We prove analogous results for the corresponding QROMs and diagonal block-encodings; cf. \Cref{theorem:be_analytic}. This reduction in depth comes at the cost of a superpolynomial, but subexponential, number of ancilla qubits.

\smallskip

\subsection{Organization of the paper} In \Cref{sec:blockencoding}, we present a method to construct diagonal block-encodings using QROMs and a (shallow) block-encoding of the position operator. In \Cref{sec:quantumstateprep}, we show how such block-encodings of diagonal operators can be used for quantum state preparation. In \Cref{sec:whqrom} we construct shallow QROMs for functions with sparse Walsh--Hadamard Transforms. In \Cref{sec:wht_for_mvp}, we find approximations of polynomial functions with sparse Walsh--Hadamard Transforms. Finally, in \Cref{sec:wh_analytic}, we combine the various results of the paper to construct our data-loading primitives for analytic functions via piecewise Taylor-polynomial approximations.

\smallskip



\subsection{AI Disclosure} We used ChatGPT and Claude for feedback, editing, and literature search. All proofs were generated and verified by the authors. The authors assume full responsibility for the correctness of the results.

\bigskip

\section{Block encoding}
\label{sec:blockencoding}

Throughout this paper, we use two identifications of binary vectors with real numbers and vectors: If $y \in \F_2^m$, then we view it either as an unsigned rational number, $\widetilde{y}$, or a signed one, $\overline{y}$, via
\begin{align}
    \widetilde{y}       &\eqdef \sum\limits_{a = 0}^{m - 1} y_a 2^{- a - 1} \in \( \left[ 0, 2^m \) \cap \Z \) / 2^m \subset \left[ 0 , 1 \right], \\
    \overline{y}    &\eqdef - y_0 + \sum\limits_{a = 1}^{m - 1} y_a 2^{- a} \in \( \left[ - 2^{m - 1}, 2^{m - 1} \) \cap \Z \) / 2^{m - 1} \subset \left[ - 1 , 1 \right].
\end{align}
Similarly, if $\mathbf{y} \eqdef \( y_0, y_1, \ldots, y_{D - 1} \) \in \( \F_2^m \)^D$ is now a vector of binary vectors, $y_a \in \F_2^m$, then
\begin{align}
    \mathbf{\widetilde{y}}   &\eqdef \( \widetilde{y}_0, \widetilde{y}_1, \ldots, \widetilde{y}_{D - 1} \) \in \( \left[ 0, 2^m \) \cap \Z \)^D / 2^m \subset \left[ 0 , 1 \right]^D, \\
    \mathbf{\overline{y}}    &\eqdef \( \overline{y}_0, \overline{y}_1, \ldots, \overline{y}_{D - 1} \) \in \( \left[ - 2^{m - 1}, 2^{m - 1} \) \cap \Z \)^D / 2^{m - 1} \subset \left[ - 1 , 1 \right]^D.
\end{align}

\medskip

We first construct a block-encoding of the position operator $Q \ket{y}_d \eqdef \overline{y} \ket{y}_d$. A similar method was used, for example, in \cite{o2025quantum}. Let $C$ be the comparator circuit
\begin{equation}
    C \ket{y}_d \ket{z}_d \eqdef (- 1)^{\overline{y} \leqslant \overline{z}} \ket{y}_d \ket{z}_d,
\end{equation}
where $(- 1)^{\overline{y} \leqslant \overline{z}} = - 1$ if $\overline{y} \leqslant \overline{z}$ and $1$ otherwise. The oracle $C$ can be constructed with $O \( \log_2 (d) \)$ depth and $O (d)$ size; cf. \cite{vandaele2026asymptotically}. Let $\textsc{be} \( Q \) \eqdef \( \id_d \otimes H^{\otimes d} \) \circ C \circ \( \id_d \otimes H^{\otimes d} \)$. Then
\begin{align}
    \textsc{be} \( Q \) \ket{y}_d \ket{0}_d &= \frac{1}{2^d} \sum\limits_{z, z^\prime \in \F_2^d} (- 1)^{\overline{y} \leqslant \overline{z}} (- 1)^{z \odot z^\prime} \ket{y}_d \ket{z^\prime}_d = \overline{y} \ket{y}_d \ket{0}_d + \ket{y}_d \ket{\mathrm{garbage}_y}_d,
\end{align}
where
\begin{equation}
    \ket{\mathrm{garbage}_y}_d \eqdef \frac{1}{2^d} \sum\limits_{z \in \F_2^d} (- 1)^{\overline{y} \leqslant \overline{z}} \sum\limits_{z^\prime \in \F_2^d - \{ 0 \}} (- 1)^{z \odot z^\prime} \ket{z^\prime}_d
\end{equation}
is an (unnormalised) vector orthogonal to $\ket{0}_d$, and where we used that the coefficient of $\ket{y}_d \ket{0}_d$ is
\begin{equation}
    \frac{1}{2^d} \sum\limits_{z \in \F_2^d} (- 1)^{\overline{y} \leqslant \overline{z}} = \frac{2^{d - 1} \( 1 + \overline{y} \) - 2^{d - 1} \( 1 - \overline{y} \)}{2^d} = \overline{y}.
\end{equation}
Thus $\( \id_d \otimes \bra{0}_d \) \textsc{be} \( Q \) \( \id_d \otimes \ket{0}_d \) = Q$, that is, $\textsc{be} \( Q \)$ is a block-encoding of $Q$.

\smallskip

Now let $f : \F_2^n \rightarrow \F_2^d$, let $\overline{f} (x) \eqdef \overline{f (x)}$, and let $A_f \ket{x}_n \eqdef \overline{f} (x) \ket{x}_n$. Note that $A_f$ is Hermitian and $\| A_f \| \leqslant 1$, so it can be block-encoded without rescaling. Assume that we have access to the QROM
\begin{equation}
    \QROM_f \ket{x}_n \ket{y}_d \eqdef \ket{x}_n \ket{y \oplus f (x)}_d,
\end{equation}
and let
\begin{equation}
    \textsc{be} \( A_f \) \eqdef \( \QROM_f^\dagger \otimes \id_d \) \circ \( \id_n \otimes \textsc{be} \( Q \) \) \circ \( \QROM_f \otimes \id_d \). \label{eq:beaf}
\end{equation}
Applying the three factors in turn gives
\begin{align}
    \ket{x}_n \ket{0}_d \ket{0}_d   &\mapsto \ket{x}_n \ket{f (x)}_d \ket{0}_d \\
        &\mapsto \ket{x}_n \( \overline{f} (x) \ket{f (x)}_d \ket{0}_d + \ket{f (x)}_d \ket{\mathrm{garbage}_{f (x)}}_d \) \\
        &\mapsto \overline{f} (x) \ket{x}_n \ket{0}_d \ket{0}_d + \ket{x}_n \ket{0}_d \ket{\mathrm{garbage}_{f (x)}}_d.
\end{align}
Since $\ket{\mathrm{garbage}_{f (x)}}_d$ is orthogonal to $\ket{0}_d$, we get $\( \id_n \otimes \bra{0}_{2d} \) \textsc{be} \( A_f \) \( \id_n \otimes \ket{0}_{2d} \) = A_f$, that is, $\textsc{be} \( A_f \)$ is a block-encoding of $A_f$, using one call each to $\QROM_f$ and $\QROM_f^\dagger$, and a further $O (d)$ Clifford + $T$ gates and $O (d)$ ancillas. Moreover, since $\textsc{be} \( Q \)$ is Hermitian (it is a product of Hermitian gates conjugated by Hadamards), so is $\textsc{be} \( A_f \)$.

\bigskip

\section{Quantum State Preparation using block-encoding of diagonal operators}
\label{sec:quantumstateprep}

Let $f :\F_2^n \rightarrow \F_2^d \cong \( \left[ - 2^{d - 1}, 2^{d - 1} \) \cap \Z \) / 2^{d - 1} \subset \left[ - 1 , 1 \right]$, as in \Cref{sec:blockencoding}, and define
\begin{equation}
    \ket{f} \eqdef \frac{1}{\| f \|_2} \sum\limits_{x \in \F_2^n} \overline{f} (x) \ket{x}_n, \quad \mbox{where } \| f \|_2 \eqdef \sqrt{\sum_{x \in \F_2^n} \overline{f} (x)^2}.
\end{equation}
Our goal now is to show that $\ket{f}$ can be prepared, up to an $\epsilon > 0$ error, using only $O \( \tfrac{2^{\nicefrac{n}{2}}}{\| f \|_2}\)$ calls to the block encoding $\textsc{be} \( A_f \)$ in \cref{eq:beaf}.

\smallskip

First, note that
\begin{align}
    \textsc{be} \( A_f \) \circ \( H^{\otimes n} \otimes \id_d \otimes \id_d \) \ket{0}_n \ket{0}_d \ket{0}_d  &= \frac{1}{2^{\nicefrac{n}{2}}} \sum\limits_{x \in \F_2^n} \ket{x}_n \( \overline{f} (x) \ket{0}_d \ket{0}_d + \ket{0}_d \ket{\mathrm{garbage}_{f (x)}}_d \)
    \\
    &= \frac{\| f \|_2}{2^{\nicefrac{n}{2}}} \ket{f} \ket{0}_d \ket{0}_d + \sqrt{1 - \frac{\| f \|_2^2}{2^n}} \ket{f^\perp}.
\end{align}
Let
\begin{align}
    k \eqdef \left\lceil \frac{\pi}{4} \frac{2^{\nicefrac{n}{2}}}{\| f \|_2} \right\rceil, \qquad
    \theta  \eqdef 2 \arcsin \( \frac{2^{\nicefrac{n}{2}}}{\| f \|_2} \sin \( \frac{\pi}{2 (2 k + 1)} \) \),
\end{align}
and
\begin{align}
    \ket{s} &\eqdef \( \( \textsc{be} \( A_f \) \otimes R_Y \( \theta \) \) \circ \( H^{\otimes n} \otimes \id_d \otimes \id_d \otimes \id_1 \right) \) \ket{0}_n \ket{0}_d \ket{0}_d \ket{0}_1, \\
    \ket{t} &\eqdef \ket{f} \ket{0}_d \ket{0}_d \ket{1}.
\end{align}
Now $\braket{s | t} = \tfrac{\| f \|_2}{2^{\nicefrac{n}{2}}} \sin \( \tfrac{\theta}{2} \) = \sin \( \frac{\pi}{2 (2 k + 1)} \) > 0$. Thus, $k$ rounds of Quantum Amplitude Amplification (QAA) prepares $\ket{t}$ perfectly, assuming perfect preparation for $\ket{s}$ and marker oracles. One can reflect about $\ket{s}$, since it can be prepared via $\textsc{be} \( A_f \) \otimes R_Y \( \theta \)$. Similarly, $\ket{t}$ can be marked using a multi-controlled $Z$ gate, controlled by the last $d + 1$ ancillas, where the control state is $\ket{0}_d \ket{1}$. Thus, to implement QAA in the Clifford + $T$ gate set, we need to implement $Y$-rotations and multi-controlled $Z$ gates with $O \( \max \( \{ n, d \} \) \)$ controls. The former, approximated to within $\frac{\epsilon}{(2k + 1)}$ for each of its $2k + 1$ uses, can be done with $O \( \log_2 \( \frac{k}{\epsilon} \) \)$ gates each and the latter can be done with $O \( \max \( \{ n, d \} \) \)$ ancillas, using $O \( \max \( \{ n, d \} \) \)$ gates, and with depth $O \( \log_2 \( \max \( \{ n, d \} \) \) \)$; cf. \cite{claudon2024polylogarithmic}. In fact, when $k$ is large and we allow $\epsilon$ error, then one can use randomized compilation to reduce the $T$-count of these gates to $O \( \log_2 \( \nicefrac{k}{\epsilon} \) \)$; cf. \cite{gosset2025multiqubittoffoliexponentiallyfewer}*{Theorem~1}.

\begin{remark}
    \label{remark:rescaling}
    If the values of $\overline{f}$ can be chosen, as is the case in \Cref{sec:wh_analytic}, then the rotation $R_Y \( \theta \)$ can be avoided. Indeed, replacing $\overline{f}$ by $c \overline{f}$, where
    \begin{equation}
        c \eqdef \frac{2^{\nicefrac{n}{2}}}{\| f \|_2} \sin \( \frac{\pi}{2 (2k + 1)} \) \leqslant 1,
    \end{equation}
    does not change $\ket{f}$, but makes the amplitude of $\ket{f} \ket{0}_d \ket{0}_d$ in $\textsc{be} \( A_{cf} \) \circ \( H^{\otimes n} \otimes \id_d \otimes \id_d \) \ket{0}_n \ket{0}_d \ket{0}_d$ exactly $\sin \( \frac{\pi}{2 (2k + 1)} \)$. Thus $k$ rounds of Quantum Amplitude Amplification \cite{brassard2002quantum} prepare $\ket{f}$ without the flag qubit, that is, with $\theta = \pi$, and no rotation needs to be synthesised. If $c \overline{f}$ is rounded to $d$ bits, the amplitude changes by at most $2^{- (d - 1)}$, and the resulting error in the prepared state is $O \( k 2^{- d} + \frac{2^{\nicefrac{n}{2}}}{\| f \|_2} 2^{- d} \)$.
\end{remark}

\bigskip

\section{Shallow QROM for functions with sparse Walsh--Hadamard Transforms}
\label{sec:whqrom}

The construction below is a version of Oracle $\mathtt{O}_2$ from \cite{nagy_novel_2024} with improved depth. 

\smallskip

Let $f :\F_2^n \rightarrow \F_2^d \cong \( \left[ - 2^{d - 1}, 2^{d - 1} \) \cap \Z \) / 2^{d - 1} \subset \left[ - 1 , 1 \right]$, as in \Cref{sec:blockencoding}. The Walsh--Hadamard Transform of $f$ is
\begin{equation}
    \widehat{f} : \F_2^n \rightarrow \rl; z \mapsto \frac{1}{2^n} \sum\limits_{x \in \F_2^n} (- 1)^{x \odot z} \overline{f} (x).
\end{equation}
Note that $\left| \widehat{f} \right| \leqslant 1$ and the values of $\widehat{f}$ can be represented on $\( n + d \)$-bits. Let
\begin{align}
    \supp \( \widehat{f} \) \eqdef \left\{ z \in \F_2^n : \widehat{f} (z) \neq 0 \right\}, \qquad
    \cW_f                   \eqdef \left| \supp \( \widehat{f} \) \right|.
\end{align}

\smallskip

For the next theorem, define our \emph{resource state} as
\begin{equation}
    \ket{\cR_f} \eqdef \bigotimes\limits_{z \in \supp \( \widehat{f} \)} \ket{\widehat{f} (z)}_{n + d},
\end{equation}
This is an $\( n + d \) \cW_f$-qubit state that can be prepared with at most $\( n + d \) \cW_f$ $X$ gates.

\smallskip

\begin{theorem}
    \label{theorem:wh_qrom}
    The oracle
    \begin{equation}
        \QROM_f \ket{x}_n \ket{0}_d \eqdef \ket{x}_n \ket{f (x)}_d,
    \end{equation}
    can be synthesised in the Clifford + $T$ gate set using:
    \begin{itemize}
        \item $O \( \( n + d \) \cW_f \log_2 \( \cW_f \) \)$ $T$ gates.
        \item $O((n+d)\cW_f)$ ancillas.
        \item A single reusable copy of $\ket{\cR_f}$.
    \end{itemize}
    Finally, the depth of $\QROM_f$ can be made $O \( \log_2 \( \( n + d \) \cW_f \) \)$.
\end{theorem}

\begin{proof}
    Let us assume that the system starts in the state $\ket{x}_n \ket{0}_d \ket{0}_{n + d} \ket{\cR_f}$, for some $x \in \F_2^n$, where $\ket{0}_d$ is the output register and $\ket{0}_{n + d}$ is a work register. We define the following gate, which acts only on the resource state: 
    \begin{equation}
        \mathrm{sign}_f \ket{x}_n \ket{0}_d \ket{0}_{n + d} \ket{\cR_f} \eqdef \ket{x}_n \ket{0}_d \ket{0}_{n + d} \bigotimes\limits_{z \in \supp \( \widehat{f} \)} \( \mathrm{add}_{2^{- \( n + d - 1 \)}} \circ X^{\otimes (n + d)} \)^{x \odot z} \ket{\widehat{f} (z)}_{n + d}.
    \end{equation}
    Note that flipping all bits of an $\( n + d \)$-bit signed rational $a$ yields $- a - 2^{- \( n + d - 1 \)}$, \footnote{The most negative value, $- 1$, has no representable negation. However, $\widehat{f} (z) = - 1$ is only possible for $z = 0$, since otherwise $\overline{f} (x) = 1$ for some $x$, which is not representable, and $z = 0$ is never negated, since $x \odot 0 = 0$.} that is,
    \begin{equation}
        X^{\otimes (n + d)} \ket{a}_{n + d} = \ket{- a - 2^{- \( n + d - 1 \)}}_{n + d},
    \end{equation}
    and thus
    \begin{equation}
        \( \mathrm{add}_{2^{- \( n + d - 1
         \)}} \circ X^{\otimes (n + d)} \)^{x \odot z} \ket{\widehat{f} (z)}_{n + d} = \ket{(- 1)^{x \odot z} \widehat{f} (z)}_{n + d}.
    \end{equation}
    Using \cite{nagy_novel_2024}*{Appendix~C~and~D}, we see that the bit-flip part of $\mathrm{sign}_f$ can be implemented with $O \( \( n + d \) \cW_f \)$ $\CNOT$ gates with depth of $O \( \log_2 \( \( n + d \) \cW_f \) \)$. The addition, for each $\( n + d \)$-qubit register, is controlled by the same bit, $x \odot z$, and can be implemented, in parallel for each such register, with $O \( \( n + d \) \cW_f \)$ Clifford + $T$ gates and with depth $O \( \log_2 \( n + d \) \)$. For any
    \begin{equation}
        g : \supp \( \widehat{f} \) \rightarrow \F_2^{n + d} \cong \left[ 0, 2^{n + d} \) \cap \Z,
    \end{equation}
    let $G \eqdef \sum_z g(z) \mod 2^{n + d}$ and $G = \sum_{a = 0}^{n + d - 1} G_a 2^{n + d - 1 - a}$. Now we define the summation gate, $\mathrm{sum}$, so that
    \begin{equation}
        \ket{0}_{n + d} \bigotimes\limits_{z \in \supp \( \widehat{f} \)} \ket{g(z)}_{n + d} \eqdef \( \bigotimes\limits_{a = 0}^{n + d - 1} \ket{G_a} \) \ket{\mathrm{garbage}_g}.
    \end{equation}
    We synthesise $\mathrm{sum}$ by recalling the $3$-to-$2$ carry-save adder: for any $n + d$-digit unsigned integers $a, b, c$ we have that $a + b + c = 2 q + r$, where $q$ is the bitwise majority and $r$ is the bitwise sum (modulo $2$) of $a, b, c$. This step can be implemented with a single temporary AND gate and one further ancilla; cf. \cite{gidney2018halving}*{Figure~4/(b)}. Thus, we can reduce the addition of the $\cW_f$ integers to the addition of $O (1)$ (but at least $2$) integers with $O \( \log_2 \( \cW_f \) \)$ layers of constant depth. Note that the number of digits needed to represent these numbers grows but it is bounded by $O \( n + d \)$, thus, one can sum these $O (1)$ numbers with depth $\log_2 \( n + d \)$ and $O \( n + d \)$ $T$ gates, using a logarithmic-depth adder circuits, such as \cite{draper2004logarithmic}. The total number of ancillas needed for this procedure is $O((n+d)\cW_f)$. This can be adjusted to work for signed integers as well, and thus, for signed rationals. Now we have that
    \begin{equation}
        \mathrm{sum} \circ \mathrm{sign}_f \ket{x}_n \ket{0}_d \ket{0}_{n + d} \ket{\cR_f} = \ket{x}_n \ket{0}_d \ket{f (x)}_d \ket{0}_n \ket{\mathrm{garbage}_{f, x}},
    \end{equation}
    since $\overline{f} (x)$ is a multiple of $2^{- (d - 1)}$, and thus the last $n$ bits of its $(n + d)$-bit representation in the work register vanish. Finally, we copy the first $d$ bits of the work register into the output register using $d$ $\CNOT$ gates, and then apply $\( \mathrm{sum} \circ \mathrm{sign}_f \)^\dagger$. This maps the state to $\ket{x}_n \ket{f (x)}_d \ket{0}_{n + d} \ket{\cR_f}$, which implements $\QROM_f$ and returns $\ket{\cR_f}$ unchanged, so it can be reused. This at most doubles the gate count and depth.
\end{proof}

\smallskip

\begin{remark}
    The proof above works verbatim if $\supp \( \widehat{f} \)$ is replaced by any set $S \supseteq \supp \( \widehat{f} \)$, with a fixed ordering, in which case $\ket{\cR_f} \eqdef \bigotimes_{z \in S} \ket{\widehat{f} (z)}_{n + d}$ contains $\ket{0}_{n + d}$ for each $z \in S - \supp \( \widehat{f} \)$.
\end{remark}

\smallskip

\begin{remark}
    The number of work ancillas can be reduced at the cost of increased depth. Partition $\supp \( \widehat{f} \)$ into $B$ blocks, compute the partial sums $\sum_z (- 1)^{x \odot z} \widehat{f} (z)$ over each block one block at a time, reusing the same work registers, and accumulate them into the output register. This uses work registers for only $O \( \frac{\cW_f}{B} \)$ summands at a time, at the cost of depth $O \( B \log_2 \( \frac{(n + d) \cW_f}{B} \) \)$. The resource state $\ket{\cR_f}$ is unaffected.
\end{remark}

\bigskip

\section{Approximate Walsh--Hadamard Transform of multivariable polynomials}
\label{sec:wht_for_mvp}

Let $\Phi$ be a real polynomial in $D$ variables, defined by
\begin{equation}
    \Phi \( t_0, t_1, \ldots, t_{D - 1} \) = \sum\limits_{\substack{\alpha \in \N^D \\ \| \alpha \|_1 \leqslant \deg \( \Phi \)}} \upphi_\alpha t^\alpha.
\end{equation}
Rescale $\Phi$ so that it maps $[0, 1]^D$ into $\left[ - \frac{1}{2}, \frac{1}{2} \right]$; this leaves room for the rounding errors below, so that all values stay in $[- 1, 1)$ for $d \geqslant 2$. For each $m \in \Z_+$, define pseudo-boolean functions as
\begin{equation}
    \Phi_m : \( \F_2^m \)^D \rightarrow \left[ - \frac{1}{2}, \frac{1}{2} \right]; \ \mathbf{y} \mapsto \Phi \( \mathbf{\widetilde{y}} \).
\end{equation}

\medskip

In this section, we show that the Walsh--Hadamard Transform of $\Phi_m$ is supported on vectors of Hamming weight at most $\deg \( \Phi \)$, and that rounding its coefficients yields a nearby function with the same sparse support. Here and below, $\mathbf{y} \odot \mathbf{v} \in \F_2$ denotes the dot product modulo $2$, $h \( \mathbf{v} \)$ denotes the Hamming weight of $\mathbf{v}$, and the inverse Walsh--Hadamard Transform is given by $\wh^{- 1} (g) \( \mathbf{y} \) \eqdef \sum_{\mathbf{v}} (- 1)^{\mathbf{y} \odot \mathbf{v}} g \( \mathbf{v} \)$. First, let us compute the Walsh--Hadamard Transform of $\Phi_m$ at $\mathbf{v} = \( v_0, v_1, \ldots, v_{D - 1} \) \in \( \F_2^m \)^D$:
\begin{align}
    \wh \( \Phi_m \) \( \mathbf{v} \)   &= \frac{1}{2^{Dm}} \sum\limits_{\mathbf{y} \in \( \F_2^m \)^D} (- 1)^{\mathbf{y} \odot \mathbf{v}} \Phi_m \( \mathbf{y} \) \\
        &= \frac{1}{2^{Dm}} \sum\limits_{\mathbf{y} \in \( \F_2^m \)^D} (- 1)^{\mathbf{y} \odot \mathbf{v}} \sum\limits_{\substack{\alpha \in \N^D \\ \| \alpha \|_1 \leqslant \deg \( \Phi \)}} \upphi_\alpha \mathbf{\widetilde{y}}^\alpha \\
        &= \sum\limits_{\substack{\alpha \in \N^D \\ \| \alpha \|_1 \leqslant \deg \( \Phi \)}} \upphi_\alpha \( \frac{1}{2^{Dm}} \sum\limits_{\mathbf{y} \in \( \F_2^m \)^D} (- 1)^{\mathbf{y} \odot \mathbf{v}} \mathbf{\widetilde{y}}^\alpha \) \\
        &= \sum\limits_{\substack{\alpha \in \N^D \\ \| \alpha \|_1 \leqslant \deg \( \Phi \)}} \upphi_\alpha \prod\limits_{a = 0}^{D - 1} \( \frac{1}{2^m} \sum\limits_{y_a \in \F_2^m} (- 1)^{y_a \odot v_a} \widetilde{y}_a^{\alpha_a} \).
\end{align}
Note that
\begin{align}
    \frac{1}{2^m} \sum\limits_{y_a \in \F_2^m} (- 1)^{y_a \odot v_a} \widetilde{y}_a^{\alpha_a} &= \frac{1}{2^m} \sum\limits_{y_a \in \F_2^m} (- 1)^{y_a \odot v_a} \( \sum\limits_{b = 0}^{m - 1} y_{a, b} 2^{- b - 1} \)^{\alpha_a} \\
        &= \frac{1}{2^m} \sum\limits_{y_a \in \F_2^m} (- 1)^{y_a \odot v_a} \sum\limits_{\substack{\beta \in \N^m \\ \| \beta \|_1 = \alpha_a}} \binom{\alpha_a}{\beta} \prod\limits_{\substack{b = 0, \ldots, m - 1 \\ \beta_b \neq 0}} \frac{y_{a, b}}{2^{(b + 1) \beta_b}} \\
        &= \sum\limits_{\substack{\beta \in \N^m \\ \| \beta \|_1 = \alpha_a}} \binom{\alpha_a}{\beta} 2^{- \sum_{b = 0}^{m - 1} (b + 1) \beta_b} \( \frac{1}{2^m} \sum\limits_{y_a \in \F_2^m} (- 1)^{y_a \odot v_a} \prod\limits_{\substack{b = 0, \ldots, m - 1 \\ \beta_b \neq 0}} y_{a, b} \).
\end{align}
We say that $v_a \in \F_2^m$ is compatible with $\beta \in \N^m$ if
\begin{equation}
    \forall b \in \{ 0, 1, \ldots, m - 1 \}: \: \beta_b = 0 \Rightarrow v_{a, b} = 0.
\end{equation}
If $v_a$ is compatible with $\beta$, then we write $v_a \subseteq \beta$. Let $|\beta| \in \Z_+$ be the number of nonzero components of $\beta$ (the cardinality of its support). With this convention, we have
\begin{equation}
    \frac{1}{2^m} \sum\limits_{y_a \in \F_2^m} (- 1)^{y_a \odot v_a} \prod\limits_{\substack{b = 0, \ldots, m - 1 \\ \beta_b \neq 0}} y_{a, b} = \left\{ \begin{array}{ll} (- 1)^{h \( v_a \)} 2^{- |\beta|}, & \mbox{if } v_a \subseteq \beta, \\ 0, & \mbox{otherwise.}  \end{array} \right.
\end{equation}
Thus,
\begin{equation}
    \wh \( \Phi_m \) \( \mathbf{v} \) = (- 1)^{h \( \mathbf{v} \)} \sum\limits_{\substack{\alpha \in \N^D \\ \| \alpha \|_1 \leqslant \deg \( \Phi \)}} \upphi_\alpha \prod\limits_{a = 0}^{D - 1} \sum\limits_{\substack{\beta \in \N^m \\ \| \beta \|_1 = \alpha_a \\ v_a \subseteq \beta}} \binom{\alpha_a}{\beta} 2^{- |\beta| - \sum_{b = 0}^{m - 1} (b + 1) \beta_b}.
\end{equation}
Let us define
\begin{equation}
    C_{\alpha_a} \( v_a \) \eqdef \sum\limits_{\substack{\beta \in \N^m \\ \| \beta \|_1 = \alpha_a \\ v_a \subseteq \beta}} \binom{\alpha_a}{\beta} 2^{- |\beta| - \sum_{b = 0}^{m - 1} (b + 1) \beta_b} \geqslant 0.
\end{equation}
Let us extend the above definition of compatibility to $v = \( v_0, v_1, \ldots, v_{D - 1} \) \in \( \F_2^m \)^D$ and $\beta = \( \beta_0, \beta_1, \ldots, \beta_{D - 1} \) \in \( \N^m \)^D$ by defining $v \subseteq \beta$ exactly when for all $a \in \{ 0, 1, \ldots, D - 1 \}$, we have $v_a \subseteq \beta_a$. Now we have
\begin{equation}
    \wh \( \Phi_m \) \( \mathbf{v} \) = (- 1)^{h \( \mathbf{v} \)} \sum\limits_{\substack{\alpha \in \N^D \\ \| \alpha \|_1 \leqslant \deg \( \Phi \)}} \upphi_\alpha \prod\limits_{a = 0}^{D - 1} C_{\alpha_a} \( v_a \). \label{eq:wh_of_poly}
\end{equation}
Note that $\mathbf{v} \subseteq \beta$ implies that $h \( \mathbf{v} \) \leqslant |\beta| \leqslant \| \alpha \|_1 \leqslant \deg \( \Phi \)$. In particular, if $\deg \( \Phi \) < \tfrac{Dm}{2}$, then the number of nonzero Walsh--Hadamard components, $\cW_\Phi$, satisfies
\begin{equation}
    \log_2 \( \cW_\Phi \) \leqslant H \( \frac{\deg \( \Phi \)}{Dm} \) Dm,
\end{equation}
where $H$ is the \emph{binary entropy}, $H(t)$, of $0 < t < 1$, defined as
\begin{equation}
    H(t) \eqdef - t \log_2 (t) - (1 - t) \log_2 (1 - t).
\end{equation}
This follows since every $\mathbf{v} \in \( \F_2^m \)^D$ with $\wh \( \Phi_m \) \( \mathbf{v} \) \neq 0$ satisfies $h \( \mathbf{v} \) \leqslant \deg \( \Phi \)$, and hence
\begin{equation}
    \cW_\Phi \leqslant \sum\limits_{w = 0}^{\deg \( \Phi \)} \binom{Dm}{w} \leqslant 2^{H \( \frac{\deg \( \Phi \)}{Dm} \) Dm}.
\end{equation}
Finally, let $\Psi_{m, d} : \( \F_2^m \)^D \rightarrow \F_2^{d + Dm}$ be the $(d + Dm)$-bit truncation of $\wh \( \Phi_m \)$, that is, for all $\mathbf{v} \in \( \F_2^m \)^D$, we have
\begin{equation}
    \overline{\Psi}_{m, d} \( \mathbf{v} \) = \left\lfloor 2^{d + Dm - 1} \wh \( \Phi_m \) \( \mathbf{v} \) \right\rfloor / 2^{d + Dm - 1} \in \( \left[ - 2^{d + Dm - 1}, 2^{d + Dm - 1} \) \cap \Z \) / 2^{d + Dm - 1} \cong \F_2^{d + Dm}.
\end{equation}
Note that
\begin{align}
    \Phi_m \( \mathbf{y} \) - \wh^{- 1} \( \overline{\Psi}_{m, d} \) \( \mathbf{y} \)  &= \sum\limits_{\mathbf{v} \in \( \F_2^m \)^D} (- 1)^{\mathbf{y} \odot \mathbf{v}} \( \wh \( \Phi_m \) \( \mathbf{v} \) - \overline{\Psi}_{m, d} \( \mathbf{v} \) \) \\
        &= \sum\limits_{\mathbf{v} \in \( \F_2^m \)^D} (- 1)^{\mathbf{y} \odot \mathbf{v}} \( \wh \( \Phi_m \) \( \mathbf{v} \) - \left\lfloor 2^{d + Dm - 1} \wh \( \Phi_m \) \( \mathbf{v} \) \right\rfloor / 2^{d + Dm - 1} \) \\
        &= \sum\limits_{\mathbf{v} \in \( \F_2^m \)^D} (- 1)^{\mathbf{y} \odot \mathbf{v}} \( 2^{d + Dm - 1} \wh \( \Phi_m \) \( \mathbf{v} \) - \left\lfloor 2^{d + Dm - 1} \wh \( \Phi_m \) \( \mathbf{v} \) \right\rfloor \) / 2^{d + Dm - 1},
\end{align}
thus,
\begin{equation}
    \left\| \Phi_m - \wh^{- 1} \( \overline{\Psi}_{m, d} \) \right\|_{\ell^\infty \( \( \F_2^m \)^D \)} \leqslant 2^{- (d - 1)}.
\end{equation}
Since each $\overline{\Psi}_{m, d} \( \mathbf{v} \)$ is a multiple of $2^{- (d + Dm - 1)}$, so is $\wh^{- 1} \( \overline{\Psi}_{m, d} \)$. Thus we can define $\Phi_{m, d} : \( \F_2^m \)^D \rightarrow \F_2^{d + Dm}$ via $\overline{\Phi}_{m, d} \eqdef \wh^{- 1} \( \overline{\Psi}_{m, d} \)$, so that the Walsh--Hadamard Transform of $\overline{\Phi}_{m, d}$ is exactly $\overline{\Psi}_{m, d}$, and hence has at most $\cW_\Phi$ nonzero components. By the result above,
\begin{equation}
    \left\| \Phi_m - \overline{\Phi}_{m, d} \right\|_{\ell^\infty \( \( \F_2^m \)^D \)} \leqslant 2^{- (d - 1)}.
\end{equation}

\smallskip

Now the following result is a corollary of our definitions and \Cref{theorem:wh_qrom}, applied to $\Phi_{m, d}$ with $n = Dm$ and with $d$ replaced by $d + Dm$.

\begin{corollary}
    \label{corollary:poly_qrom}
    The oracle
    \begin{equation}
        \QROM_{\Phi, m, d} \ket{\mathbf{y}}_{Dm} \ket{0}_{d + Dm} \eqdef \ket{\mathbf{y}}_{Dm} \ket{\Phi_{m, d} \( \mathbf{y} \)}_{d + Dm},
    \end{equation}
    can be synthesised in the Clifford + $T$ gate set using:
    \begin{itemize}
        \item $O \( \( Dm + d \) H \( \frac{\deg \( \Phi \)}{Dm} \) Dm \, 2^{H \( \frac{\deg \( \Phi \)}{Dm} \) Dm} \)$ $T$ gates.
        \item $O \( \( Dm + d \) 2^{H \( \frac{\deg \( \Phi \)}{Dm} \) Dm} \)$ ancillas, in addition to the resource state.
        \item A single reusable copy of $\ket{\cR_{\Phi_{m, d}}}$.
    \end{itemize}
    Finally, the depth of $\QROM_{\Phi, m, d}$ can be made $O \( \log_2 \( Dm + d \) + H \( \frac{\deg \( \Phi \)}{Dm} \) Dm \)$.
\end{corollary}

\bigskip

\section{Shallow Oracles for analytic functions}
\label{sec:wh_analytic}

In this section, we consider functions that are analytic in a neighborhood of $[0, 1]^D$ and construct shallow QROMs for such functions, which immediately give rise to shallow block-encodings and quantum state preparation circuits via the results of \Cref{sec:blockencoding,sec:quantumstateprep}.

\smallskip

In order to understand the appropriate error rate encoding smooth, multivariable functions, we prove the following lemma, motivated by \cite{zylberman2025efficient}*{Theorem IV.1} and \cite{szczepanik2026simulating}*{Theorem A.1}.

\begin{lemma}
    \label{lemma:smooth_encoding_scaling}
    Let $\Phi : [0, 1]^D \rightarrow [- 1, 1]$ be Lipschitz continuous with Lipschitz constant $K_\Phi$, and not identically zero. For each $m \in \Z_+$ and $\mathbf{x} \in \( \F_2^m \)^D$, let us define
    \begin{align}
        A_{\Phi, m} \ket{\mathbf{x}}    \eqdef \Phi \( \widetilde{\mathbf{x}}\) \ket{\mathbf{x}}, \qquad
        \ket{\Phi, m} \eqdef \frac{1}{N_{\Phi, m}} \sum\limits_{\mathbf{x} \in \( \F_2^m \)^D} \Phi \( \widetilde{\mathbf{x}}\) \ket{\mathbf{x}},
    \end{align}
    where $N_{\Phi, m}^2 \eqdef \sum\limits_{\mathbf{x} \in \( \F_2^m \)^D} \left| \Phi \( \widetilde{\mathbf{x}}\) \right|^2$. Then, for all $m, m^\prime \in \Z_+$, with $m \leqslant m^\prime$, we have
    \begin{subequations}
    \begin{align}
        \left\| A_{\Phi, m^\prime} - A_{\Phi, m} \otimes \id_{D \( m^\prime - m \)} \right\|                &= O_\Phi \( 2^{- m} \), \label{eq:aphi_asym} \\
        \left\| \ket{\Phi, m^\prime} - \ket{\Phi, m} \ket{+}^{\otimes D \( m^\prime - m \)} \right\|  &= O_\Phi \( 2^{- m} \), \label{eq:ketphi_asym}
    \end{align}
    \end{subequations}
    where $\| \cdot \|$ denotes the operator norm or the Hilbert norm.
\end{lemma}

\begin{proof}
    Let $l \eqdef m^\prime - m \in \Z_+$. For all $\( \mathbf{x}, \mathbf{y} \) \in \( \F_2^m \oplus \F_2^l \)^D$, we have
    \begin{equation}
        \Phi \( \widetilde{\( \mathbf{x}, \mathbf{y} \)} \) - \Phi \( \widetilde{\mathbf{x}} \) = \Phi \( \widetilde{\mathbf{x}} + 2^{- m} \widetilde{\mathbf{y}} \) - \Phi \( \widetilde{\mathbf{x}} \) = O_\Phi \( 2^{- m} \).
    \end{equation}
    which proves \cref{eq:aphi_asym}.

        Next, consider the vectors $u$ and $w$, with one entry for each $\( \mathbf{x}, \mathbf{y} \) \in \( \F_2^m \oplus \F_2^l \)^D$, given by
    \begin{equation}
        u \( \mathbf{x}, \mathbf{y} \) \eqdef \Phi \( \widetilde{\mathbf{x}} + 2^{- m} \widetilde{\mathbf{y}} \), \qquad w \( \mathbf{x}, \mathbf{y} \) \eqdef \Phi \( \widetilde{\mathbf{x}} \).
    \end{equation}
    Thus $u$ samples $\Phi$ on the finer grid, while $w$ repeats the value at $\widetilde{\mathbf{x}}$ for all $\mathbf{y}$. Since $\| u \| = N_{\Phi, m^\prime}$ and $\| w \| = 2^{\nicefrac{Dl}{2}} N_{\Phi, m}$, we have $\ket{\Phi, m^\prime} = \frac{u}{\| u \|}$ and $\ket{\Phi, m} \ket{+}^{\otimes Dl} = \frac{w}{\| w \|}$. By the triangle inequality and $\left| \| w \| - \| u \| \right| \leqslant \| u - w \|$,
    \begin{equation}
        \left\| \ket{\Phi, m^\prime} - \ket{\Phi, m} \ket{+}^{\otimes Dl} \right\| \leqslant \frac{\| u - w \|}{\| u \|} + \| w \| \left| \frac{1}{\| u \|} - \frac{1}{\| w \|} \right| \leqslant \frac{2 \| u - w \|}{\| u \|}.
    \end{equation}
    Since $\widetilde{\mathbf{y}} \in [0, 1)^D$, we have $\left| 2^{- m} \widetilde{\mathbf{y}} \right| \leqslant \sqrt{D} 2^{- m}$, so, by the Lipschitz continuity of $\Phi$, each of the $2^{D (m + l)}$ entries of $u - w$ has absolute value at most $\sqrt{D} K_\Phi 2^{- m}$. Hence
    \begin{equation}
        \| u - w \| \leqslant \sqrt{D} K_\Phi 2^{- m} 2^{\nicefrac{D (m + l)}{2}}.
    \end{equation}
    Moreover, $2^{- D (m + l)} \| u \|^2$ is a Riemann sum for $\int_{[0, 1]^D} \left| \Phi \( \mathbf{x} \) \right|^2 d^D \mathbf{x}$, which is positive since $\Phi$ is continuous and not identically zero. Thus $\| u \| \geqslant c_\Phi 2^{\nicefrac{D (m + l)}{2}}$ for some $c_\Phi > 0$ and all $m$ large enough. Combining these bounds, the factors of $2^{\nicefrac{D (m + l)}{2}}$ cancel, and
    \begin{equation}
        \left\| \ket{\Phi, m^\prime} - \ket{\Phi, m} \ket{+}^{\otimes Dl} \right\| \leqslant \frac{2 \sqrt{D} K_\Phi}{c_\Phi} 2^{- m} = O_\Phi \( 2^{- m} \),
    \end{equation}
    which proves \cref{eq:ketphi_asym}.
\end{proof}

\smallskip

\begin{remark}
    The results of \Cref{lemma:smooth_encoding_scaling} can be interpreted as the admissible error asymptotics for encoding smooth data; an approximate construction for $A_{\Phi, m}$ (resp. $\ket{\Phi, m}$) that differs by an error of $O_\Phi \( 2^{- m} \)$ in the operator norm (resp. by an error of $O_\Phi \( 2^{- m} \)$ in the Hilbert norm) asymptotically captures the same amount from the original data, $\Phi$.
    
    More informally, as $m \rightarrow \infty$, such approximations ``converge to the continuous limit'' at the same rate.
\end{remark}

The goal of this section is to give an approximate construction for $A_{\Phi, m}$ and $\ket{\Phi, m}$ that satisfy \cref{eq:aphi_asym,eq:ketphi_asym}, when $\Phi$ is analytic in a neighborhood of the hypercube.

\smallskip

Let
\begin{equation}
    \Phi : [0, 1]^D \rightarrow \left[ - \frac{1}{2}, \frac{1}{2} \right],
\end{equation}
be a function that extends analytically to a neighborhood of $[0, 1]^D$. For $q \in \N$ and $u \in [0, 1]^D$, let
\begin{equation}
    \left\| T_q \Phi_u \right\| \eqdef \sum\limits_{\substack{\alpha \in \N^D \\ \| \alpha \|_1 = q}} \frac{\left| \partial^\alpha \Phi (u) \right|}{\alpha!}
\end{equation}
be the $\ell^1$-norm of the $q^{\mathrm{th}}$ order Taylor coefficients of $\Phi$ at $u$, and for $p \in \N$ let
\begin{equation}
    \rho_p \eqdef \sup \( \left\{ \sqrt[q]{\left\| T_q \Phi_u \right\|} : u \in [0, 1]^D, \ q > p \right\} \).
\end{equation}
Clearly, $\rho_p$ is non-increasing in $p$. Moreover, $\rho_p < \infty$: by compactness, there are $r, M > 0$ such that $\Phi$ extends holomorphically to the polydisc of polyradius $r$ around each $u \in [0, 1]^D$, with $| \Phi | \leqslant M$ there, so the Cauchy estimates give $\left\| T_q \Phi_u \right\| \leqslant M \binom{q + D - 1}{D - 1} r^{- q}$, whose $q^{\mathrm{th}}$ root is bounded in $q$.

\smallskip

Let $k, m \in \Z_+$. Each coordinate is discretized with $k + m$ qubits, the first $k$ of which select a box of side length $2^{- k}$. If $\( \mathbf{x}, \mathbf{y} \) \in \( \F_2^k \)^D \oplus \( \F_2^m \)^D \cong \( \F_2^k \oplus \F_2^m \)^D$, then $\widetilde{\( \mathbf{x}, \mathbf{y} \)} = \widetilde{\mathbf{x}} + 2^{- k} \widetilde{\mathbf{y}}$. Assume that $\rho_0 2^{- k} \leqslant \frac{1}{2}$. Since $\widetilde{\mathbf{y}} \in [0, 1)^D$, we have $\widetilde{\mathbf{y}}^\alpha \leqslant 1$, and hence the Taylor series
\begin{equation}
    \Phi_{k, m} \( \mathbf{x}, \mathbf{y} \) = \Phi \( \widetilde{\mathbf{x}} + 2^{- k} \widetilde{\mathbf{y}} \) = \sum\limits_{\alpha \in \N^D} \frac{\partial^\alpha \Phi \( \widetilde{\mathbf{x}} \)}{\alpha!} \frac{\widetilde{\mathbf{y}}^\alpha}{2^{k \| \alpha \|_1}}
\end{equation}
converges absolutely, since its terms with $\| \alpha \|_1 = q$ contribute at most $\left\| T_q \Phi_{\widetilde{\mathbf{x}}} \right\| 2^{- kq} \leqslant \( \rho_0 2^{- k} \)^q$, and it equals $\Phi$ on the box by analyticity. Similarly, truncating it to $\| \alpha \|_1 \leqslant p$ leaves a remainder of absolute value at most
\begin{equation}
    \sum\limits_{q = p + 1}^\infty \( \rho_p 2^{- k} \)^q \leqslant 2 \( \rho_p 2^{- k} \)^{p + 1}. \label{eq:taylor_remainder}
\end{equation}
We use this observation below for our constructions.

Let us compute the Walsh--Hadamard Transform of $\Phi_{k,m}$, but only in the second variable; using \cref{eq:wh_of_poly}, we get
\begin{align}
    \wh_2 \( \Phi_{k, m} \) \( \mathbf{x}, \mathbf{v} \)  &= (- 1)^{h \( \mathbf{v} \)} \sum\limits_{\alpha \in \N^D} \frac{\partial^\alpha \Phi \( \widetilde{\mathbf{x}} \)}{\alpha!} \frac{\wh \( \widetilde{\mathbf{y}}^\alpha \) \( \mathbf{v} \)}{2^{k \| \alpha \|_1}} \\
        &= (- 1)^{h \( \mathbf{v} \)} \sum\limits_{\alpha \in \N^D} \frac{\partial^\alpha \Phi \( \widetilde{\mathbf{x}} \)}{\alpha! 2^{k \| \alpha \|_1}} \prod\limits_{a = 0}^{D - 1} C_{\alpha_a} \( v_a \) \\
        &= (- 1)^{h \( \mathbf{v} \)} \sum\limits_{\substack{\alpha \in \N^D \\ \| \alpha \|_1 \geqslant h \( \mathbf{v} \)}} \frac{\partial^\alpha \Phi \( \widetilde{\mathbf{x}} \)}{\alpha! 2^{k \| \alpha \|_1}} \prod\limits_{a = 0}^{D - 1} C_{\alpha_a} \( v_a \).
\end{align}
Define $\Phi_{k, m, p}^{(\infty)}$ so that
\begin{equation}
    \wh_2 \( \Phi_{k, m, p}^{(\infty)} \) \( \mathbf{x}, \mathbf{v} \) \eqdef (- 1)^{h \( \mathbf{v} \)} \sum\limits_{\substack{\alpha \in \N^D \\ p \geqslant \| \alpha \|_1 \geqslant h \( \mathbf{v} \)}} \frac{\partial^\alpha \Phi \( \widetilde{\mathbf{x}} \)}{\alpha! 2^{k \| \alpha \|_1}} \prod\limits_{a = 0}^{D - 1} C_{\alpha_a} \( v_a \).
\end{equation}
Let $S_p \eqdef \left\{ \mathbf{v} \in \F_2^{Dm} : h \( \mathbf{v} \) \leqslant p \right\}$, with a fixed ordering, so that the Walsh--Hadamard Transform of $\Phi_{k, m, p}^{(\infty)} \( \mathbf{x}, \cdot \)$ is supported in $S_p$ for every $\mathbf{x}$. In particular, $\wh_2 \( \Phi_{k, m, p}^{(\infty)} \) \( \mathbf{x}, \mathbf{v} \) = 0$ whenever $h \( \mathbf{v} \) > p$. Now, by \cref{eq:taylor_remainder}, we have
\begin{equation}
    \left| \Phi_{k, m} \( \mathbf{x}, \mathbf{y} \) - \Phi_{k, m, p}^{(\infty)} \( \mathbf{x}, \mathbf{y} \) \right| \leqslant 2 \( \rho_p 2^{- k} \)^{p + 1}.
\end{equation}
By \cref{eq:aphi_asym} in \Cref{lemma:smooth_encoding_scaling}, $A_{\Phi_{k, m, p}^{(\infty)}, k + m}$ is an asymptotically good approximation of $A_{\Phi, k + m}$, if
\begin{equation}
    \left\| \Phi_{k, m} - \Phi_{k, m, p}^{(\infty)} \right\|_{\ell^\infty \( \( \F_2^k \oplus \F_2^m \)^D \)} = O_\Phi \( 2^{- (k + m)} \).
\end{equation}
Hence, we need
\begin{equation}
    k + m \leqslant (p + 1) k - (p + 1) \log_2 \( \rho_p \) + O_\Phi (1).
\end{equation}
which implies
\begin{equation}
    m \leqslant kp - (p + 1) \log_2 \( \rho_p \) + O_\Phi (1).
\end{equation}
Note that this inequality holds for
\begin{equation}
    p_m \eqdef \left\lceil \sqrt{\frac{m}{\log_2 \( D m \)}} \right\rceil \quad \& \quad k_m \eqdef \max \( \left\{ 1, \left\lceil \sqrt{m \log_2 \( D m \)} + \( 1 + \frac{1}{p_m} \) \log_2 \( \rho_{p_m} \) \right\rceil \right\} \). \label{eq:kp_def}
\end{equation}
Indeed, since $p_m \geqslant \sqrt{\nicefrac{m}{\log_2 (Dm)}}$, we have $k_m p_m \geqslant p_m \sqrt{m \log_2 (Dm)} + \( p_m + 1 \) \log_2 \( \rho_{p_m} \) \geqslant m + \( p_m + 1 \) \log_2 \( \rho_{p_m} \)$. Moreover, since $k_m \rightarrow \infty$, the assumption $\rho_0 2^{- k_m} \leqslant \frac{1}{2}$ holds for all sufficiently large $m$.

\smallskip

Next, for each $\mathbf{x} \in \F_2^{Dk}$, let 
\begin{equation}
    \Phi_{k, p, m}^\mathbf{x} \( \mathbf{y} \) \eqdef \sum\limits_{\substack{\alpha \in \N^D \\ \| \alpha \|_1 \leqslant p}} \frac{\partial^\alpha \Phi \( \widetilde{\mathbf{x}} \)}{\alpha!} \frac{\widetilde{\mathbf{y}}^\alpha}{2^{k \| \alpha \|_1}},
\end{equation}
and let $\Phi_{k, p, m, d}^\mathbf{x} : \F_2^{Dm} \rightarrow \F_2^{d + Dm}$ be defined, as in \Cref{sec:wht_for_mvp}, by rounding the Walsh--Hadamard coefficients of $\Phi_{k, p, m}^\mathbf{x}$ down to $d + Dm$ bits and inverting exactly. Then the Walsh--Hadamard Transform of $\overline{\Phi}_{k, p, m, d}^\mathbf{x}$ is supported in $S_p$, and $\left\| \Phi_{k, p, m}^\mathbf{x} - \overline{\Phi}_{k, p, m, d}^\mathbf{x} \right\|_{\ell^\infty} \leqslant 2^{- (d - 1)}$. Using \Cref{corollary:poly_qrom}, we can construct $\QROM_{\Phi_{k, p, m, d}^\mathbf{x}}$ using only Clifford and $T$ gates, using a single reusable copy of $\ket{\cR_{\Phi_{k, p, m, d}^\mathbf{x}}}$ and with a depth of $O \( \log_2 \( Dm + d \) + H \( \frac{p}{Dm} \) Dm \)$. Now we construct the approximation of $A_{\Phi_{k, m, p}^{(\infty)}, k + m}$ as follows:
\begin{enumerate}
    \item Let us assume that we start in the basis state $\ket{\mathbf{x}}_{Dk} \ket{\mathbf{y}}_{Dm} \ket{0}_{d + Dm} \ket{0}_{d + Dm}$.
    \item Prepare ancillas in the state $\bigotimes\limits_{\mathbf{x}^\prime \in \F_2^{Dk}} \ket{\cR_{\Phi_{k, p, m, d}^{\mathbf{x}^\prime}}}$, where all resource states are indexed by the same ordered set $S_p$, padded with zeros. This is an $O \( \( 2 Dm + d \) 2^{Dk + H \( \frac{p}{Dm} \) Dm} \)$-qubit state, but can be prepared with a single layer of $X$ gates.
    \item Use the Swap QROM from \cites{low_trading_2024,motlagh2026halving} to bring $\ket{\cR_{\Phi_{k, p, m, d}^\mathbf{x}}}$ to the beginning of the ancilla register. The depth of this step is $O \( Dk \)$.
    \item Since all resource states have the same layout, the remaining part of the construction of $\QROM_{\Phi_{k, p, m, d}^\mathbf{x}}$ is independent of $\mathbf{x}$; hence, load $\Phi_{k, p, m, d}^\mathbf{x} \( \mathbf{y} \)$:
    \begin{equation}
        \ket{\mathbf{x}}_{Dk} \ket{\mathbf{y}}_{Dm} \ket{0}_{d + Dm} \ket{0}_{d + Dm} \mapsto \ket{\mathbf{x}}_{Dk} \ket{\mathbf{y}}_{Dm} \ket{\Phi_{k, p, m, d}^\mathbf{x} \( \mathbf{y} \)}_{d + Dm} \ket{0}_{d + Dm}.
    \end{equation}
    The depth of this step is $O \( \log_2 \( Dm + d \) + H \( \frac{p}{Dm} \) Dm \)$.
    \item Apply the block-encoding of the $(d + Dm)$-bit position operator from \Cref{sec:blockencoding} to the last two registers. The depth of this step is $O \( \log_2 \( d + Dm \) \)$.
    \item Uncompute by applying the inverses of steps 4 and 3, which returns the ancillas to $\bigotimes_{\mathbf{x}^\prime} \ket{\cR_{\Phi_{k, p, m, d}^{\mathbf{x}^\prime}}}$.
\end{enumerate}
Since $H (t) \leqslant t \log_2 \( \nicefrac{e}{t} \)$ for $t \in \left( 0, \frac{1}{2} \right]$, we have $H \( \frac{p_m}{Dm} \) Dm \leqslant p_m \log_2 \( e D m \) = O \( \sqrt{m \log_2 \( D m \)} \)$. The total depth is $O \( Dk + \log_2 \( Dm + d \) + H \( \frac{p}{Dm} \) Dm \)$.

\smallskip

Now we are ready to state the first main theorem of the section.

\begin{theorem}
    \label{theorem:be_analytic}
    Let $\Phi : [0, 1]^D \rightarrow \left[ - \frac{1}{2}, \frac{1}{2} \right]$ extend analytically to a neighborhood of $[0, 1]^D$, and let $p_m$ and $k_m$ be as in \cref{eq:kp_def}. Then, for all sufficiently large $m$ and all $d \in \Z_+$, there exists a block-encoding of $A_{\Phi, k_m + m}$, up to an error of $O_\Phi \( 2^{- \min \( \left\{ k_m + m, d - 1 \right\} \)} \)$ in the operator norm, in the Clifford + $T$ gate set, with depth
    \begin{equation}
        O \( D k_m + \log_2 \( Dm + d \) + H \( \frac{p_m}{Dm} \) Dm \),
    \end{equation}
    using $O \( \( 2 Dm + d \) 2^{D k_m + H \( \frac{p_m}{Dm} \) Dm} \)$ ancilla qubits. In particular, if $d = O(m)$ and $D$ is fixed, then the depth is $O \( \sqrt{m \log_2 \( D m \)} \)$.
\end{theorem}

\smallskip

\begin{remark}
    When $\Phi$ extends to an analytic function on all of $\cx^D$, then $\lim_{p \rightarrow \infty} \rho_p = 0$, so $\log_2 \( \rho_{p_m} \) \rightarrow - \infty$, and $k_m$, and hence the depth, can be smaller, depending on the decay rate of $\rho_p$.

    When $\Phi$ only extends to an analytic function on a neighborhood of $[0, 1]^D$, then $\log_2 \( \rho_{p_m} \)$ is bounded, and thus the depth is $O \( \sqrt{m \log_2 \( D m \)} \)$, where the implied constant depends on $\Phi$ only through $\rho_0$.
\end{remark}

\smallskip

Let us turn to the problem of approximate state preparation for $\ket{\Phi, k + m}$. In this case, the relevant gauge of error is the Hilbert norm. Fortunately, the Walsh--Hadamard Transform is unitary, up to constant factor. In this case, we truncate the Walsh--Hadamard Transform by Hamming weight. Let $p \in \Z_+$ and define $\Phi_{k, m, p}^{(2)}$ so that
\begin{equation}
    \wh_2 \( \Phi_{k, m, p}^{(2)} \) \( \mathbf{x}, \mathbf{v} \) \eqdef \left\{ \begin{array}{ll} \wh_2 \( \Phi_{k, m} \) \( \mathbf{x}, \mathbf{v} \), & \mbox{if } h \( \mathbf{v} \) \leqslant p, \\ 0, & \mbox{otherwise,} \end{array} \right.
\end{equation}
where we recall that
\begin{equation}
    \wh_2 \( \Phi_{k, m} \) \( \mathbf{x}, \mathbf{v} \) = (- 1)^{h \( \mathbf{v} \)} \sum\limits_{\substack{\alpha \in \N^D \\ \| \alpha \|_1 \geqslant h \( \mathbf{v} \)}} \frac{\partial^\alpha \Phi \( \widetilde{\mathbf{x}} \)}{\alpha! 2^{k \| \alpha \|_1}} \prod\limits_{a = 0}^{D - 1} C_{\alpha_a} \( v_a \).
\end{equation}
Next, we compute
\begin{align}
    \left\| \Phi_{k, m} - \Phi_{k, m, p}^{(2)} \right\|_{\ell^2 \( \( \F_2^k \oplus \F_2^m \)^D \)}^2 &= \sum\limits_{\mathbf{x} \in \F_2^{Dk}} \sum\limits_{\substack{\mathbf{v} \in \F_2^{Dm} \\ h \( \mathbf{v} \) > p}} 2^{Dm} \left| \wh_2 \( \Phi_{k, m} \) \( \mathbf{x}, \mathbf{v} \) \right|^2 \\
        &= 2^{Dm} \sum\limits_{\mathbf{x} \in \F_2^{Dk}} \sum\limits_{\substack{\mathbf{v} \in \F_2^{Dm} \\ h \( \mathbf{v} \) > p}} \left| (- 1)^{h \( \mathbf{v} \)} \sum\limits_{\substack{\alpha \in \N^D \\ \| \alpha \|_1 \geqslant h \( \mathbf{v} \)}} \frac{\partial^\alpha \Phi \( \widetilde{\mathbf{x}} \)}{\alpha! 2^{k \| \alpha \|_1}} \prod\limits_{a = 0}^{D - 1} C_{\alpha_a} \( v_a \) \right|^2 \\
        &= 2^{Dm} \sum\limits_{\mathbf{x} \in \F_2^{Dk}} \sum\limits_{\substack{\mathbf{v} \in \F_2^{Dm} \\ h \( \mathbf{v} \) > p}} \left| \sum\limits_{w = h \( \mathbf{v} \)}^\infty \frac{1}{2^{k w}} \sum\limits_{\substack{\alpha \in \N^D \\ \| \alpha \|_1 = w}} \frac{\partial^\alpha \Phi \( \widetilde{\mathbf{x}} \)}{\alpha!} \prod\limits_{a = 0}^{D - 1} C_{\alpha_a} \( v_a \) \right|^2.
\end{align}
We now bound the inner sum. Since $|\beta| \geqslant h \( v_a \)$ whenever $v_a \subseteq \beta$, and by the multinomial theorem, we have
\begin{equation}
    C_{\alpha_a} \( v_a \) \leqslant 2^{- h \( v_a \)} \sum\limits_{\substack{\beta \in \N^m \\ \| \beta \|_1 = \alpha_a}} \binom{\alpha_a}{\beta} 2^{- \sum_{b = 0}^{m - 1} (b + 1) \beta_b} = 2^{- h \( v_a \)} \( 1 - 2^{- m} \)^{\alpha_a} \leqslant 2^{- h \( v_a \)},
\end{equation}
and hence, by the definition of $\rho_p$,
\begin{equation}
    \left| \sum\limits_{\substack{\alpha \in \N^D \\ \| \alpha \|_1 = w}} \frac{\partial^\alpha \Phi \( \widetilde{\mathbf{x}} \)}{\alpha!} \prod\limits_{a = 0}^{D - 1} C_{\alpha_a} \( v_a \) \right| \leqslant 2^{- h \( \mathbf{v} \)} \sum\limits_{\substack{\alpha \in \N^D \\ \| \alpha \|_1 = w}} \frac{\left| \partial^\alpha \Phi \( \widetilde{\mathbf{x}} \) \right|}{\alpha!} \leqslant 2^{- h \( \mathbf{v} \)} \rho_{w - 1}^w.
\end{equation}
Since $\rho_p$ is non-increasing in $p$ and $\rho_0 2^{- k} \leqslant \frac{1}{2}$, for $h \( \mathbf{v} \) > p$ we get
\begin{equation}
    \left| \wh_2 \( \Phi_{k, m} \) \( \mathbf{x}, \mathbf{v} \) \right| \leqslant 2^{- h \( \mathbf{v} \)} \sum\limits_{w = h \( \mathbf{v} \)}^\infty \( \rho_p 2^{- k} \)^w \leqslant 2 \( \rho_p 2^{- (k + 1)} \)^{h \( \mathbf{v} \)}.
\end{equation}
Let $q \eqdef \rho_p 2^{- (k + 1)}$. Since there are $\binom{Dm}{h}$ vectors $\mathbf{v} \in \F_2^{Dm}$ with $h \( \mathbf{v} \) = h$,
\begin{equation}
    \left\| \Phi_{k, m} - \Phi_{k, m, p}^{(2)} \right\|_{\ell^2 \( \( \F_2^k \oplus \F_2^m \)^D \)}^2 \leqslant 4 \cdot 2^{D (k + m)} \sum\limits_{h = p + 1}^{Dm} \binom{Dm}{h} q^{2h}.
\end{equation}
For $m$ large enough, we have $Dm q^2 \leqslant \frac{1}{2}$, so consecutive terms of the last sum decrease at least by a factor of $2$, and thus, using $\binom{Dm}{p + 1} \leqslant 2^{H \( \frac{p + 1}{Dm} \) Dm}$,
\begin{equation}
    \left\| \Phi_{k, m} - \Phi_{k, m, p}^{(2)} \right\|_{\ell^2 \( \( \F_2^k \oplus \F_2^m \)^D \)}^2 \leqslant 8 \cdot 2^{D (k + m) + H \( \frac{p + 1}{Dm} \) Dm} \( \rho_p 2^{- (k + 1)} \)^{2 (p + 1)}.
\end{equation}
Since
\begin{equation}
    \| \Phi_{k, m} \|_{\ell^2 \( \( \F_2^k \oplus \F_2^m \)^D \)}^2 = \Theta \( 2^{D (k + m)} \left\| \Phi \right\|_{L^2 \( [0, 1]^D \)}^2 \) = \Theta_\Phi \( 2^{D (k + m)} \),
\end{equation}
and
\begin{align}
    \left\| \ket{\Phi, k + m} - \frac{\Phi_{k, m, p}^{(2)}}{\| \Phi_{k, m, p}^{(2)} \|_2} \right\| &= \left\| \frac{\Phi_{k, m}}{\| \Phi_{k, m} \|_{\ell^2 \( \( \F_2^k \oplus \F_2^m \)^D \)}} - \frac{\Phi_{k, m, p}^{(2)}}{\| \Phi_{k, m, p}^{(2)} \|_{\ell^2 \( \( \F_2^k \oplus \F_2^m \)^D \)}} \right\|_{\ell^2 \( \( \F_2^k \oplus \F_2^m \)^D \)} \\
        &= O \( \frac{\left\| \Phi_{k, m} - \Phi_{k, m, p}^{(2)} \right\|_{\ell^2 \( \( \F_2^k \oplus \F_2^m \)^D \)}}{\| \Phi_{k, m} \|_{\ell^2 \( \( \F_2^k \oplus \F_2^m \)^D \)}} \) \\
        &= O \( \left\| \Phi \right\|_{L^2 \( [0, 1]^D \)}^{- 1} \rho_p^{p + 1} 2^{\frac{1}{2} H \( \frac{p + 1}{Dm} \) Dm - (k + 1) (p + 1)} \).
\end{align}
By \cref{eq:ketphi_asym}, $\frac{\Phi_{k, m, p}^{(2)}}{\| \Phi_{k, m, p}^{(2)} \|_2}$ is an asymptotically good approximation of $\ket{\Phi, k + m}$ when
\begin{equation}
    \left\| \ket{\Phi, k + m} - \frac{\Phi_{k, m, p}^{(2)}}{\| \Phi_{k, m, p}^{(2)} \|_2} \right\| = O_\Phi \( 2^{- (k + m)} \).
\end{equation}
Hence, we need
\begin{equation}
    k + m \leqslant (k + 1)(p + 1) - \frac{1}{2} H \( \frac{p + 1}{Dm} \) Dm - (p + 1) \log_2 \( \rho_p \) + O_\Phi (1). \label{eq:k_p_m_relation}
\end{equation}
This holds for $p = p_m$ and
\begin{equation}
    k = k_m^{(2)} \eqdef k_m + \left\lceil \log_2 \( e D m \) \right\rceil. \label{eq:k2_def}
\end{equation}
Indeed, writing $E \eqdef \left\lceil \log_2 \( e D m \) \right\rceil$, we have $k_m p_m \geqslant m + \( p_m + 1 \) \log_2 \( \rho_{p_m} \)$ by \cref{eq:kp_def}, and $H \( \frac{p_m + 1}{Dm} \) Dm \leqslant \( p_m + 1 \) \log_2 \( e D m \) \leqslant \( p_m + 1 \) E \leqslant 2 p_m E$, since $H (t) \leqslant t \log_2 \( \nicefrac{e}{t} \)$ for $t \in \left( 0, \frac{1}{2} \right]$. Thus the right-hand side of \cref{eq:k_p_m_relation} exceeds its left-hand side by at least $p_m + 1$. Since $k_m^{(2)} - k_m = O \( \log_2 \( Dm \) \)$, this does not change the order of the depth.

\smallskip

Next we study the depth of the corresponding quantum state preparation circuit. As in the construction of the block-encoding above, for each $\mathbf{x} \in \F_2^{Dk}$ we round the Walsh--Hadamard coefficients $\wh_2 \( \Phi_{k, m, p}^{(2)} \) \( \mathbf{x}, \cdot \)$ down to $d + Dm$ bits and invert exactly. Let $f \eqdef \Phi_{k, m, p, d}^{(2)} : \F_2^{D (k + m)} \rightarrow \F_2^{d + Dm}$ be the resulting function. Its Walsh--Hadamard Transform in the second variable is supported in $S_p$ for every $\mathbf{x}$, and, as in \Cref{sec:wht_for_mvp}, $\left\| \Phi_{k, m, p}^{(2)} - \overline{f} \right\|_{\ell^\infty} \leqslant 2^{- (d - 1)}$. In particular,
\begin{equation}
    \left\| \Phi_{k, m} - \overline{f} \right\|_{\ell^2 \( \( \F_2^k \oplus \F_2^m \)^D \)} \leqslant \left\| \Phi_{k, m} - \Phi_{k, m, p}^{(2)} \right\|_{\ell^2 \( \( \F_2^k \oplus \F_2^m \)^D \)} + 2^{\nicefrac{D (k + m)}{2}} 2^{- (d - 1)},
\end{equation}
so, for $m$ and $d$ large enough, $\| f \|_2 = \Theta \( 2^{\nicefrac{D (k + m)}{2}} \| \Phi \|_{L^2 \( [0, 1]^D \)} \)$, and $\ket{f}$ is within $O_\Phi \( 2^{- (k + m)} + 2^{- (d - 1)} \)$ of $\ket{\Phi, k + m}$.

By \Cref{sec:quantumstateprep}, using the rescaling of \Cref{remark:rescaling} to avoid synthesising $R_Y \( \theta \)$, the state $\ket{f}$ can be prepared using $O \( \frac{2^{\nicefrac{D (k + m)}{2}}}{\| f \|_2} \) = O \( \| \Phi \|_{L^2 \( [0, 1]^D \)}^{- 1} \) = O_\Phi (1)$ calls to $\textsc{be} \( A_f \)$, together with reflections of depth $O \( \log_2 \( D (k + m) + d \) \)$. Since the Walsh--Hadamard Transform of $f$ in the second variable is supported in $S_p$ for every $\mathbf{x}$, the block-encoding of $A_f$ is constructed exactly as that of $A_{\Phi_{k, m, p, d}^{(\infty)}}$ above, with depth
\begin{equation}
    O \( Dk + \log_2 \( Dm + d \) + H \( \frac{p}{Dm} \) Dm \).
\end{equation}

\begin{theorem}
    \label{theorem:sp_analytic}
    Let $\Phi : [0, 1]^D \rightarrow \left[ - \frac{1}{2}, \frac{1}{2} \right]$ extend analytically to a neighborhood of $[0, 1]^D$ and be not identically zero, and let $p_m$ and $k_m^{(2)}$ be as in \cref{eq:kp_def,eq:k2_def}. Then, for all sufficiently large $m$ and all $d \geqslant k_m^{(2)} + m + 1$, the state $\ket{\Phi, k_m^{(2)} + m}$ can be prepared, up to an error of $O_\Phi \( 2^{- \( k_m^{(2)} + m \)} \)$ in the Hilbert norm, in the Clifford + $T$ gate set, with depth
    \begin{equation}
        O_\Phi \( D k_m^{(2)} + \log_2 \( Dm + d \) + H \( \frac{p_m}{Dm} \) Dm \).
    \end{equation}
    In particular, if $d = O(m)$ and $D$ is fixed, then the depth is $O_\Phi \( \sqrt{m \log_2 \( D m \)} \)$.
\end{theorem}

\bibliography{references}

\end{document}